# Room temperature antiferromagnetic resonance and inverse spin-Hall voltage in canted antiferromagnets

I. Boventer[1], H. T. Simensen[2], A. Anane[1], M. Kläui[2,3,4], A. Brataas[2], R. Lebrun[1]

1. Unité Mixte de Physique CNRS, Thales, Univ. Paris-Sud, Université Paris-Saclay, Palaiseau 91767, France
2. Center for Quantum Spintronics, Department of Physics, Norwegian University of Science and Technology, Trondheim, Norway
3. Institut für Physik, Johannes Gutenberg-Universität Mainz, D-55099, Mainz, Germany
4. Graduate School of Excellence Materials Science in Mainz (MAINZ), Staudingerweg 9, D-55128, Mainz, Germany

**We study theoretically and experimentally the spin pumping signals induced by the resonance of canted antiferromagnets with Dzyaloshinskii-Moriya interaction and demonstrate that they can generate easily observable inverse spin-Hall voltages. Using a bilayer of hematite/heavy metal as a model system, we measure at room temperature the antiferromagnetic resonance and an associated inverse spin-Hall voltage, as large as in collinear antiferromagnets. As expected for coherent spin-pumping, we observe that the sign of the inverse spin-Hall voltage provides direct information about the mode handedness as deduced by comparing hematite, chromium oxide and the ferrimagnet Yttrium-Iron Garnet. Our results open new means to generate and detect spin-currents at terahertz frequencies by functionalizing antiferromagnets with low damping and canted moments.**

Contemporary spintronics, utilizing the electronic spin for information processing and microelectronics, is mostly based on ferromagnetic device architectures. In view of long term perspectives to enable enhanced data processing speeds and downscaling for on-chip information processing [1], spintronics with antiferromagnets is a promising avenue [2]. Antiferromagnets exhibit the key advantage over ferromagnets that their resonance frequency is enhanced by the exchange coupling of the sublattices, and thus generally in the terahertz regime [2,3]. In compensated antiferromagnets, the absence of a net moment however strongly impedes simple access to their ultrafast dynamics, especially in thin films, and the development of ultra-fast antiferromagnet-based devices [4,5]. As a result, interfacial spin-transport phenomena could provide new insights into the spin-relaxation processes and spin-dynamics in antiferromagnets [5–8].

Experimental access to the spin dynamics can be facilitated by spin to charge conversion mechanisms such as the spin pumping effect largely studied in ferromagnets [9,10]. The spin pumping effect generates alternating (ac) and continuous (dc) spin-currents from the spin precession of a magnetic material into an adjacent conductor [11,12]. These spin-currents can be detected electrically by measuring a voltage via the inverse spin-Hall effect (ISHE) [12] or through the inverse Rashba-Edelstein effect [13]. Applied to antiferromagnets (AFMs), these combined spintronic effects could provide a direct and surface sensitive access to the magnetization dynamics in both bulk AFM and thin films. However, theoretical [14,15] and recent experimental studies [16,17] showed that, in collinear antiferromagnets, the amplitude of the pumped spin-currents scales with the dynamical sub-lattice symmetry breaking. Thus, its amplitude is proportional between the ratio of the anisotropy field $H_A$ and the exchange field $H_E$. This ratio is of less than 0.1% in many compounds [5,6,15], and only reaches 1-2% in two known compounds $MnF_2$ [18] and $FeF_2$ [19] without applying large magnetic fields of the order of the spin-flop field. This limitation has until now largely restrained the investigation of spin-pumping signals in antiferromagnets.

In parallel, routes to generate spin-pumping signals from noncollinear antiferromagnets have not been widely considered yet. Early studies have reported that the bulk Dzyaloshinskii Moriya interaction (DMI) can reach a few Tesla in some antiferromagnets, and induce small canted moments both for easy axis (e.g. $NiF_2$ [20], $CoF_2$ [20] or most orthoferrites [21,22]) and easy plane antiferromagnets (e.g. $MnCO_3$ [20] or hematite, α-$Fe_2O_3$, above the Morin transition [23]), or chiral antiferromagnets (e.g. $Mn_3Sn$ [24]).

In this Letter, we explore both theoretically and experimentally spin-pumping in non-collinear antiferromagnets with a DMI induced canting and capped with a heavy metal. First, we demonstrate theoretically that the AFM resonance generates a dc spin-pumping and an inverse spin-Hall voltage $V_{ISHE}$ in the adjacent heavy metal, proportional to the ratio $H_D/H_E$ ($H_D$ : DMI field, $H_E$ : exchange field). This result is valid in both easy-axis and easy-plane canted AFMs. Note, that the zero-field mode frequencies do not depend directly on the DMI field. We anticipate inverse spin-Hall voltages $V_{ISHE} >$ 100 nV in many canted AFMs with $H_D/H_E \propto 1 - 10\ \%$ [23,25]. In parallel, their zero-field mode frequencies can range from tens to hundreds of gigahertz (depending on the exchange $H_E$ and anisotropy $H_A$ fields of the materials [23,25]). Second, we experimentally study hematite (α-Fe$_2$O$_3$) capped with platinum to confirm our theoretical predictions. Due to the low Gilbert damping and its residual anisotropy in the easy-plane phase [23,26,27], we easily measure the resonance of the low frequency mode $f_-$ of hematite [6,23,28,29] and its associated $V_{ISHE}$. We report $V_{ISHE} > 30$ nV at 300 K as large as in the uniaxial antiferromagnet Chromium oxide (Cr$_2$O$_3$) at low temperatures and only one order of magnitude smaller than in the ferrimagnetic insulator YIG. Furthermore, a direct comparison between the signs of the $V_{ISHE}$ in the three compounds allows an identification of their respective mode handedness. Altogether, our results highlight that canted antiferromagnets embrace the rich dynamics of antiferromagnets, whilst keeping a net moment as in ferromagnets, and a larger temperature stability than ferrimagnets with compensated angular momentum [30].

We start by deriving theoretically the spin-pumping response associated with the magnetization dynamics of both canted easy-axis and easy-plane antiferromagnets when put in contact with normal metals. To this end, we model the magnetization dynamics in a macrospin approximation with two sublattices. We include the contributions from a DMI field $\boldsymbol{H_D}$ and an external magnetic field $\boldsymbol{H}$ orthogonal to the Néel order parameter. Both these fields contribute to the canted moments and, hence, to the emergence of a finite magnetic moment. The system's free energy reads:

$$F = M\left[H_E(\boldsymbol{m_A} \cdot \boldsymbol{m_B}) - H_D\hat{\boldsymbol{z}} \cdot (\boldsymbol{m_A} \times \boldsymbol{m_B}) + \frac{H_A}{2}(m_{A,z}^2 + m_{B,z}^2) - \frac{H_a}{2}(m_{A,y}^2 + m_{B,y}^2) - \boldsymbol{H} \cdot (\boldsymbol{m_A} + \boldsymbol{m_B})\right] \quad (1),$$

where γ denotes the gyromagnetic ratio, $\boldsymbol{m_i}$ the unit vector of the sublattice magnetization $i$ ($i \in \{A, B\}$), $\boldsymbol{M_i} = M\boldsymbol{m_i}$, and $\boldsymbol{H} = H\hat{\boldsymbol{x}}$ is the externally applied static magnetic field, $H_A$ denotes the hard-axis anisotropy along the $z$ axis, and $H_a$ the easy-axis anisotropy within the $xy$ plane. The above model can be applied on both canted easy-axis ($H_A = 0$) and canted easy-plane ($H_A \gg H_a$) AFMs. In these AFMs, the eigenmodes are non-degenerate with a low frequency mode $f_-$ and a high frequency mode $f_+$:

$$\begin{cases} f_- = \left(\frac{\gamma}{2\pi}\right)\sqrt{2H_E H_a + H(H + H_D)} & (2) \\ f_+ = \left(\frac{\gamma}{2\pi}\right)\sqrt{2H_E(H_a + H_A) + H_D(H + H_D)} & (3) \end{cases}.$$

The eigenmodes are qualitatively similar in canted easy-plane and canted easy-axis AFMs (see Supplementary Material [31]). Hence, we can continue with a generalized model that describes both systems. This contrasts with the collinear easy-axis and easy-plane AFMs, in which the modes are profoundly different and need separate treatments [14–16,32,33]. The low frequency mode of canted AFMs is characterized by a right-handed elliptical precession of the magnetic moment (see **Fig. 1. (a)**). The high frequency mode is characterized by a linearly oscillating magnetization (see **Suppl. Mat.** [31]). Note that, in the absence of an applied field, the gap of the low frequency mode [Eq. (2)] depends only on the exchange $H_E$ and easy-axis anisotropy $H_a$, and not on the DMI field $H_D$. This remarkable feature causes a low frequency mode in the THz range in canted easy-axis AFMs such as ErFeO$_3$ [34]. In canted easy-plane AFMs where $H_a$ is often much smaller, the low frequency mode can even be found in the range of a few GHz, such as for hematite above the Morin temperature [23].

We next derive an expression for the inverse spin Hall voltage $V_{ISHE}$ induced by spin pumping from the low frequency mode into an adjacent heavy metal layer. We consider excitations by an oscillating magnetic field $\boldsymbol{h_{ac}}$ (applied along the $\boldsymbol{y}$ axis) at the resonance frequency $f_-$, which induces the precession of the magnetic moments within the (yz) plane. The transverse $V_{ISHE}$ voltage (along **y**) is proportional to the dynamic magnetization amplitudes in both the $y$ and $z$ directions. Following the approach of Ref. [14], we establish the expression for the inverse spin-Hall voltage $V_{ISHE}$ [see **Suppl. Mat.** [31] for details]:

$$V_{ISHE} = \frac{\hbar \theta_N}{8} \frac{d_V}{d_N} \gamma^3 \left(\frac{h_{ac}}{H_E}\right)^2 Q^2 \frac{(H+H_D)^3}{4\pi^2 f^2} \frac{\lambda e G_R \tan(d_N/2\lambda)}{\hbar\sigma + 2\lambda e^2 G_R \coth(d_N/\lambda)}, \quad (4)$$

where $G_R$ denotes the real part of the spin mixing conductance per unit area of the interface, $h_{ac}$ is the amplitude of the excitation field, $e$ is the elementary charge, $d_V$ is the distance between the voltage leads, $d_N$ is the Pt layer's thickness, $\theta_N$, $\lambda$, and $\sigma$ are the thickness, spin diffusion length and the conductivity of Pt, respectively. $Q_- = \frac{f_-}{\Delta f_-}$ corresponds to the material quality factor with $\Delta f_- = \alpha(\gamma/\pi)H_E$ the linewidth of the low frequency resonance peak [28]. The expression of the antiferromagnetic linewidth and the damping values remain under strong debate in antiferromagnets [23,28,35,36]. However, resonance measurements in insulating antiferromagnets generally show $Q_-$ factors in the range of 100 - 1000 [37–41]. We thus chose to evaluate the expected inverse spin-Hall voltages $V_{ISHE}$ for prototypical canted antiferromagnets such as orthoferrites with $Q_-$ factors of 500, in the range of the existing reports [37,40,41], and note that it can reach 0.5 µV (for $h_{ac}$ = 1 mT, i.e. one tenth of the ac field used in Ref. [16]). For a given $Q_-$ factor, $V_{ISHE}$ scales with $\frac{(H+H_D)^3}{H_E^2 f_-^2}$. As $f_-$ scales with $H$, we anticipate a linear increase of the inverse spin-Hall voltage $V_{ISHE}$ with the applied field (for H >> H_D) which is opposite to the ferromagnetic case [42].

| Material | $H_E$ (T) | $H_A$ (T) | $H_a$ (T) | $H_D$ (T) | $f_-$ (THz) | $V_{ISHE}$ (nV) |
|---|---|---|---|---|---|---|
| α-Fe$_2$O$_3$ [23,26,28] | 1000 | 2.10$^{-3}$ | 6.10$^{-5}$ | 2.26 | 0.022 | 200 |
| YFeO$_3$ [43,44] | 640 | / | 0.06 | 14 | 0.25 | 711 |
| ErFeO$_3$ [45,46] | 600 | / | 0.03 | 10 | 0.17 | 628 |
| TmFeO$_3$ [30,31] | 550 | / | 0.1 | 4 | 0.3 | 18 |

*Table 1: Expected inverse spin Hall effect voltages $V_{ISHE}$ with their leading quantities $H_E$, $H_A$, $H_a$ and $H_D$ (taken at room temperature) for each listed canted antiferromagnets. All voltages are calculated with a material quality factor $Q_- = 500$ for the different materials, an excitation field $h_{ac}=1$ mT, an external field H=0.2 T to ensure a mono-domainization, and a distance between the voltage leads of 3 mm. $G_R$ is set to $6 \times 10^{18}$, λ to 1.2 nm, $\theta_N$ to 0.1 taken from Refs. [26,42,48].*

The predicted inverse spin-Hall voltages presented in **Table 1** are all above 10 nV which is in the accessible range of conventional voltage measurements [12,16,17,49]. Therefore, canted antiferromagnets with DMI induced canting open very promising perspectives for examining the electrical response of antiferromagnetic dynamics in the terahertz regime.

To confirm our predictions, we next experimentally investigate the antiferromagnetic resonance and the generated inverse-spin Hall voltage in 500 µm thick bulk crystals of hematite covered [26] by a 3 nm thick Pt layer. We then compare the recorded inverse spin Hall effect voltages to the ones of the ferrimagnet YIG and the easy-axis antiferromagnet Cr$_2$O$_3$ [17] to also obtain further information such as the mode handedness as expected for a coherent spin-pumping signal.

In the antiferromagnetic insulator hematite, the DMI field induces a small canted moment (~ 3 emu/cm$^3$) of the two sub-lattices above the Morin temperature (T$_M$ ~ 250 K). For T > T$_M$, its magnetic configuration corresponds to a canted easy-plane phase with a residual in-plane anisotropy H$_a$ such that it exhibits a low frequency mode in the gigahertz (GHz) range [23,28,29]. Therefore, we can conduct our measurements using a state-of the art highly sensitive wideband resonance spectrometer (1 - 40 GHz) with a broadband coplanar waveguide (c.f. **Suppl. Mat. Fig. 1**) and at room temperature. In **Fig. 1. (a-c)**, we present the dynamics and the frequency dispersion of the low frequency mode of hematite which has a gap around 15 GHz in agreement with previous reports [23,28].

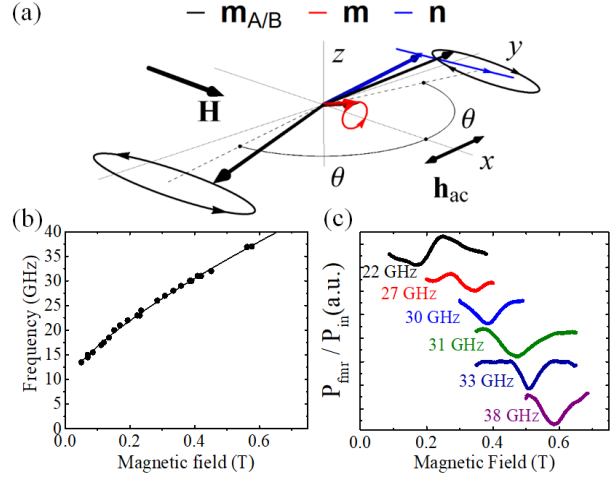

*Figure 1: Magnetic resonance of the low frequency mode of hematite. (a) Illustration of the low frequency mode of hematite in the easy plane phase in presence of a DMI induced canting. (b) The frequency dispersion of the low frequency mode of hematite measured from 10 to 40 GHz. Using Eq. (2), we extract the parameters for the anisotropy field $H_a=6\times10^{-5}$ T and $H_D=2.26$ T. (c) Resonance curves for different values of the externally applied field. The data is normalized by the input power.*

Correspondingly, we characterize the spin-pumping efficiency at the α-Fe$_2$O$_3$/Pt interface by measuring the voltage generated at resonance in the top platinum layer using a lock-in technique. This allows us to validate our model and investigate the spin dynamics of this non-collinear antiferromagnet. We measure a voltage peak at resonance only in the transverse configuration (see inset of **Fig. 2(a)**), and a clear sign reversal when we reverse the direction of the applied magnetic field. This sign inversion demonstrates the spin-pumping origin of the generated voltage. This result is to our knowledge the first evidence of spin-pumping from an easy-plane antiferromagnet and, more generally, at room temperature for an antiferromagnet.

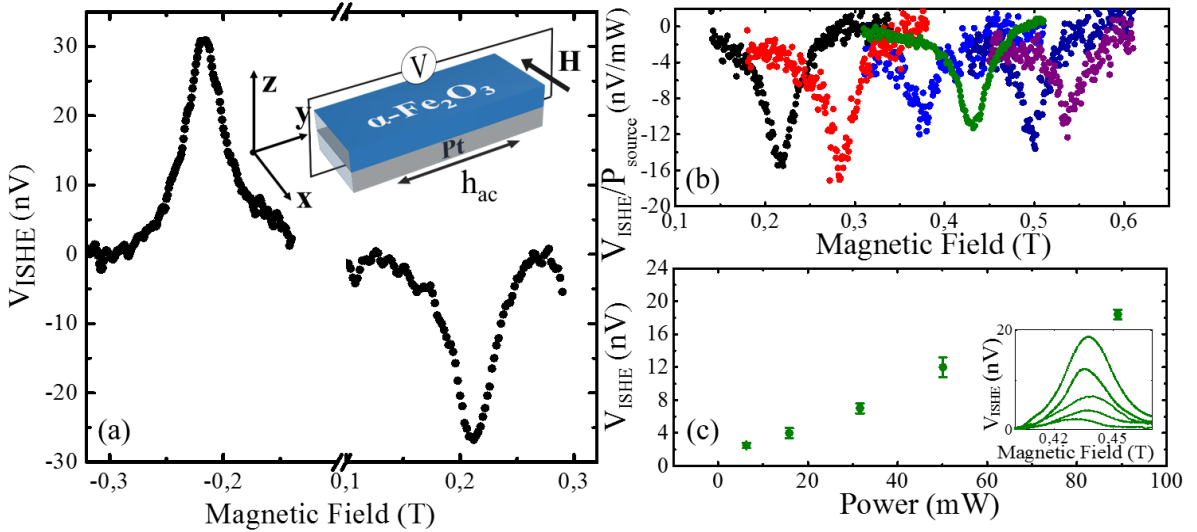

*Figure 2: Inverse spin Hall effect voltages $V_{ISHE}$ in the easy plane antiferromagnet hematite with a Dzyaloshinskii Moriya induced canted moment. (a) ISHE voltage measurement at ±0.2 T in transverse configuration (See inset for different signs of the external field **H**. The antisymmetric signal shape indicates the recorded voltage to originate from spin-pumping. (b) $V_{ISHE}$ for different frequencies as a function of the external magnetic field. Colour coding of frequencies is consistent with Fig. 1 (c). (c) Dependence of $V_{ISHE}$ peak as a function of the applied microwave power for a fixed excitation frequency of 31 GHz. In the measured range, the ISHE voltage increases linearly as expected from **Eq. (4)**. Inset shows the field dependency for the different powers.*

As for ferromagnets [12], we only measure a non-zero $V_{ISHE}$ in the transverse configuration when the magnetic field is perpendicular to the voltage contacts as shown in **Fig. 2 (a)**. In this configuration, the dc spin accumulation **σ** generated at resonance must be parallel to the applied field and thus directed along the canted net moment in order to generate a non-zero charge current $J_c \propto J_S \times \sigma$, leading to $V_{ISHE}$. This result seems at first contradictory to the spin-transport measurements in easy-plane antiferromagnets where pairs of propagating magnons carry spin-angular momentum along the direction of the antiferromagnet Néel order [29,50]. However, there is a key difference between the excitation processes. For the spin-transport case, the current induced spin-accumulation generates pairs of correlated magnon modes with different wave vectors **k**, leading to an effective non-zero spin-angular momentum along the Néel order. For the antiferromagnetic resonance case, the microwave magnetic field excites only uniform oscillations (**k**= 0) which are linearly polarized in an easy-plane system. Thus, only the dynamics of the canted moment, induced by the external magnetic field **H** or the DMI field **$H_D$**, can generate at resonance a non-zero dc spin-accumulation and thus a dc inverse spin-Hall voltage. As shown in **Fig. 2 (c)**, we also observed a linear increase of $V_{ISHE}$ with the input power, confirming that we are still in a linear regime of excitation.

Having detected the inverse spin-Hall voltage we need to determine the origin of the pumped spin current. This origin is a key point in a long-standing debate in ferro- and ferrimagnetic systems [51,52] and, more recently, also debated for antiferromagnets [16,17,53,54]. At resonance, oscillations of the excited mode [14,15] or incoherent contributions from thermal magnons (due to a resonance induced thermal gradient [55]) can both contribute to a spin-pumping signal. However, right (RH)- and left-handed (LH) circularly polarized modes carry opposite angular momentum and should thus result in inverse spin-Hall voltages with opposite signs [16,17]. Thus, one can obtain key information about the mode contributing to the spin-pumping signals by analysing the sign of the inverse spin-Hall voltages.

Using broadband coplanar waveguides up to 40 GHz, we can only access the low frequency RH mode above the Morin transition, and the LH mode below the Morin transition (see Supplemental). However, we did not detect any $V_{ISHE}$ for the LH of hematite at all temperatures below the Morin transition (See Supplemental [31]). This observation is consistent with a coherent spin-pumping signal associated to the dynamical sub-lattice symmetry breaking [14], which is extremely small in the easy-axis phase of hematite ($H_A/H_E \approx 10^{-6}$). To analyse its $V_{ISHE}$ sign, we thus measure under the same conditions the inverse spin-Hall voltages from the LH mode of the easy-axis antiferromagnet $Cr_2O_3$ [17] and the RH mode of the ferrimagnet YIG [49]. To detect the LH mode of $Cr_2O_3$, we perform measurements close to the spin-flop field to reduce the mode frequency below 40 GHz. We observe that the LH mode of $Cr_2O_3$ and the RH modes of α-$Fe_2O_3$ and YIG show inverse spin-Hall voltages with opposite signs as expected from a coherent spin-pumping model. In line with previous reports on $Cr_2O_3$ [17], the $V_{ISHE}$ of the LH mode disappears at high temperature (see Supplemental [31]) which is an indication of its coherent origin. Furthermore, the RH modes of α-$Fe_2O_3$ and YIG generate inverse spin Hall voltages with opposite signs compared to the thermal spin-pumping contribution of the RH mode of $Cr_2O_3$ reported in Ref. [17]. Lastly, the inverse spin-Hall voltages have comparable amplitudes for the LH mode of the easy-axis AFM $Cr_2O_3$ and for the RH mode of the canted AFM α-$Fe_2O_3$, and are smaller than in YIG by less than an order of magnitude. This feature indicates that both collinear and noncollinear antiferromagnets can efficiently generate spin-pumping whilst significantly enhancing the operating frequency of spintronic devices.

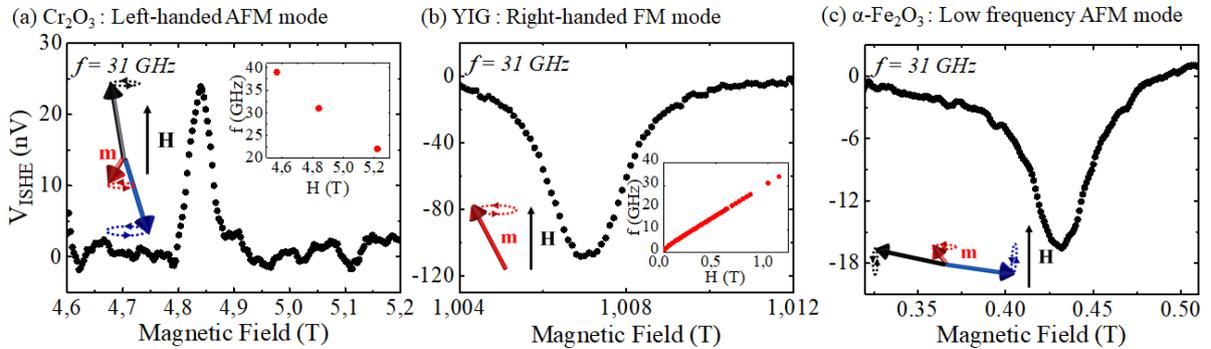

*Figure 3. Inverse spin-Hall voltages recorded at 31 GHz for (a) the left-handed AFM mode of the easy-axis antiferromagnet $Cr_2O_3$ (b) the right-handed FM mode for the ferrimagnet YIG and (c) the low frequency (right-handed) mode of the canted easy-plane antiferromagnet α-$Fe_2O_3$ capped with a platinum layer. For an inverse spin-Hall voltage originating from coherent spin pumping, a difference in handedness of the mode's polarisation is expected which induces a change in the sign of $V_{ISHE}$. We confirm the correlation between the sign of the inverse spin-Hall voltage $V_{ISHE}$ and the mode handedness since the sign changes from (a) to (b) and is equal from (b) to (c). Insets show the dispersion curves for the different modes.*

*Illustrations show the orientation of the magnetic moments with respect to the external static field H. The thickness of the YIG film is 200 nm, and the two AFMs crystals are 500 μm thick. The measurements on YIG and α-Fe$_2$O$_3$ are performed at room temperature and the ones on Cr$_2$O$_3$ are performed at 30K (see Supplemental [31]).*

In summary, we theoretically and experimentally demonstrate that noncollinear antiferromagnets with DMI induced canting can generate sizeable spin-pumping signals and associated inverse spin-Hall voltages. Using hematite as a room temperature model canted AFM, we measured a signal of $V_{ISHE} >$ 30 nV for the low frequency right-handed mode confirms our theoretical predictions. By comparing with the right-handed mode of YIG and the left-handed mode of Cr$_2$O$_3$, we confirm that the mode handedness determines the sign of the inverse spin-Hall voltage as expected for a coherent spin-pumping signal. Consequently, canted antiferromagnets allow for accessing the spin dynamics of the hitherto almost unexplored spin dynamics of noncollinear antiferromagnets. Such extensions not only broaden the understanding of the physics of the spin dynamics and of the relaxation processes for various classes of antiferromagnets but also represents a step further towards the realization of THz applications based on antiferromagnetic spintronics.

**Acknowledgement**

R.L., A.A. and M.K. acknowledge financial support from the Horizon 2020 Framework Programme of the European Commission under FET-Open grant agreement no. 863155 (s-Nebula). M.K. acknowledges support from the Graduate School of Excellence Materials Science in Mainz (MAINZ) DFG 266, the DAAD (Spintronics network, Project No. 57334897). M.K. acknowledges support from the DFG project number 423441604 and additional support from SFB TRR 173 Spin+X (projects A01 and B02 # 268565370). H.T.S., A.B., M.K. were supported by the Research Council of Norway through its Centres of Excellence funding scheme, project number 262633 "QuSpin". H.T.S. and A.B. acknowledge support from the European Research Council via Advanced Grant No. 669442 "Insulatronics".